\documentclass[aps,pra,twocolumn,groupedaddress,amsmath,superscriptaddress]{revtex4}

\usepackage{amsfonts}
\usepackage{amssymb}
\usepackage{amsmath}
\usepackage[dvips]{graphicx}
\usepackage{color}

\usepackage{graphicx}
\usepackage{hyperref}

\usepackage{mathrsfs}
\usepackage{isomath}
\usepackage{amsthm}
\usepackage{epstopdf}
\usepackage{txfonts}

\begin{document}

\title{\hspace*{-.6cm}$\mbox{Quantifying quantum coherence of optical cat states}$}
\author{Miao Zhang$^{\dag}$}
\address{The State Key Laboratory of Quantum Optics and Quantum Optics Devices,
Institute of Opto-Electronics, Shanxi University, Taiyuan 030006, China}
\author{Haijun Kang$^{\dag}$}
\address{The State Key Laboratory of Quantum Optics and Quantum Optics Devices,
Institute of Opto-Electronics, Shanxi University, Taiyuan 030006, China}
\author{Meihong Wang}
\address{The State Key Laboratory of Quantum Optics and Quantum Optics Devices,
Institute of Opto-Electronics, Shanxi University, Taiyuan 030006, China}
\address{Collaborative Innovation Center of Extreme Optics, Shanxi University, Taiyuan, Shanxi 030006, China}
\author{Fengyi Xu}
\address{The State Key Laboratory of Quantum Optics and Quantum Optics Devices,
Institute of Opto-Electronics, Shanxi University, Taiyuan 030006, China}
\author{Xiaolong Su}
\email{suxl@sxu.edu.cn}
\address{The State Key Laboratory of Quantum Optics and Quantum Optics Devices,
Institute of Opto-Electronics, Shanxi University, Taiyuan 030006, China}
\address{Collaborative Innovation Center of Extreme Optics, Shanxi University, Taiyuan, Shanxi 030006, China}
\author{Kunchi Peng}
\address{The State Key Laboratory of Quantum Optics and Quantum Optics Devices,
Institute of Opto-Electronics, Shanxi University, Taiyuan 030006, China}
\address{Collaborative Innovation Center of Extreme Optics, Shanxi University, Taiyuan, Shanxi 030006, China}

\begin{abstract}
Optical cat state plays an essential role in quantum computation and quantum metrology. Here, we experimentally quantify quantum coherence of an optical cat state by means of relative entropy and $l_{1}$ norm of coherence in Fock basis based on the prepared optical cat state at rubidium D1 line. By transmitting the optical cat state through a lossy channel, we also demonstrate the robustness of quantum coherence of optical cat state in the presence of loss, which is different from the decoherence properties of fidelity and Wigner function negativity of the optical cat state. 
Our results confirm that quantum coherence of optical cat states is robust against loss and pave the way for the application with optical cat states.
\end{abstract}

\maketitle


As a superposition of two coherent states, optical cat state is an important quantum resource for quantum information processing, including quantum computation \cite{Ralph2003, Jeong2008, Lund2008}, quantum teleportation \cite{Braunstein1998, van Enk2001, Loock2008}, and quantum metrology \cite{Gilchrist2004}. An optical cat state can be experimentally prepared by photon subtraction from a squeezed vacuum state \cite{Dakna1997, Ourjoumtsev2006, Neergaard-Nielsen2006, Wakui2007, Asavanant2017, Serikawa2018}. It has been shown that the amplitude of optical cat states can be increased by time-separated two-photon subtraction \cite{Takahashi2008} and three-photon subtraction \cite{Gerrits2010}. Meanwhile, optical squeezed cat states are generated by different means \cite{Ourjoumtsev2007, Etesse2015, Huang2015}. Based on the prepared optical cat states, several applications have been experimentally demonstrated, including quantum teleportation of cat states \cite{Lee2011}, tele-amplification \cite{Neergaard-Nielsen2013}, preparation of hybrid entangled states and teleportation based on it \cite{Jeong2014, Morin2014, Ulanov2017, Sychev2018}, amplification of cat states \cite{Lvovsky2017}, and Hadamard gate \cite{Tipsmark2011}.

Quantum coherence, which encapsulates the idea of superposition of quantum states, is a defining feature of quantum mechanics and plays a key role in applications of quantum physics and quantum information \cite{Streltsov2017}. The resource theory of quantum coherence has attracted considerable interest recently \cite{Baumgratz2014, Girolami2014, Jianwei2016, Zhang2016, Streltsov2015, Tan2018, Kaifeng2018, Matteo2019, Bagan2016}. The relations between quantum coherence and other quantum resources are extensively discussed, such as entanglement \cite{Streltsov2015, Tan2018}, asymmetry \cite{Kaifeng2018, Matteo2019} and path information \cite{Bagan2016}. Several experiments related to quantum coherence are demonstrated, including obtainment of maximal coherence via assisted distillation process \cite{Kang-Da2017}, observability and operability of robustness of coherence \cite{Zheng2018}, wave-particle duality relation based on coherence measure \cite{Yuan2018}, the trade-off relation for quantum coherence \cite{WeiMin2018}, the relation between coherence and path information \cite{Jun2018}, resilient effect of quantum coherence to transversal noise \cite{Chao2019}, the relation between quantum non-Markovianity and coherence \cite{Kang-Da2020}, experimental control of the degree of non-classicality via quantum coherence \cite{Smirne2020}, the estimation of quantum coherence by collective measurements \cite{Yuan2020}, quantification of coherence of a tunable quantum detector \cite{Huichao2020}. The roles of quantum coherence in Deutsch-Jozsa algorithm \cite{Hillery2016} and Grover quantum search algorithm \cite{Shi2017} are also discussed recently.

For application, it is crucial to quantify quantum coherence of various quantum states expressed in finite and infinite dimensional systems \cite{Baumgratz2014, Girolami2014, Jianwei2016, Zhang2016}. Very recently, K. C. Tan et al. proposed an approach to quantify the coherence between coherent states based on the Glauber-Sudarshan $P$ distribution \cite{Tan2017}, which can be used to quantify quantum coherence of the cat state. For the preparation of cat state, the density matrix in Fock basis can be obtained directly in the process of quantum tomography. Thus, it is convenient to quantify quantum coherence of cat states in Fock basis. At the same time, it has been shown that the Wigner function negativity of optical cat state is sensitive to loss \cite{Jeannic2018, Serafini2004}. However, it is unclear what's the effect of loss on quantum coherence of optical cat states. In consequence, it is essential to investigate the evolution of quantum coherence of cat states in the presence of loss.

In this paper, we prepare an optical cat state at rubidium D1 line which is a crucial medium for quantum memory. Then we experimentally demonstrate quantification of quantum coherence of the optical cat state in Fock basis by relative entropy and $l_{1}$ norm calculated from the density matrix. Finally, the evolution of quantum coherence of the optical cat state in a lossy channel is investigated. By transmitting the prepared optical cat state with amplitude of $1.06$ and fidelity of $0.68$ through a lossy channel, we show that quantum coherence of the cat state is robust against loss. Comparing with fidelity and Wigner function negativity of the cat state in a lossy channel, we show that when the fidelity is below $0.5$ and the negativity disappears, the quantum coherence of the cat state still exists. The presented results provide useful reference for the application of the optical cat state.


The relative entropy of quantum coherence refers to the distance between the quantum state $\hat \rho$ and $\hat \rho_{diag}$ formed from the diagonal elements of $\hat \rho$ \cite{Baumgratz2014}, which is given by
\begin{equation}
C_{rel.ent.}(\hat{\rho})=S(\hat{\rho}_{diag})-S(\hat{\rho}),
\label{eq:rel.ent}
\end{equation}
where $S$ is the von Neumann entropy and it is defined by $S(\hat\rho)=-$ Tr $(\hat{\rho}\log_{2}\hat{\rho})$.

The $l_{1}$ norm of quantum coherence depends on the magnitudes of off-diagonal density matrix elements \cite{Baumgratz2014}, which is given by
\begin{equation}
C_{l_{1}}(\hat{\rho})=\sum_{\substack{m,n\\ m\neq n}}{|\rho_{m,n}|}.
\label{eq:l1norm}
\end{equation}

For a cat state, generally, it is represented in infinite dimensional Fock basis. But in the experiment, the obtained density matrix of the cat state is truncated in a finite dimension by performing the quantum state tomography. Thus, the relative entropy and $l_{1}$ norm in Fock basis can be applied to quantify quantum coherence of cat states.

\begin{figure}[t]
\begin{center}
\includegraphics[width=80mm]{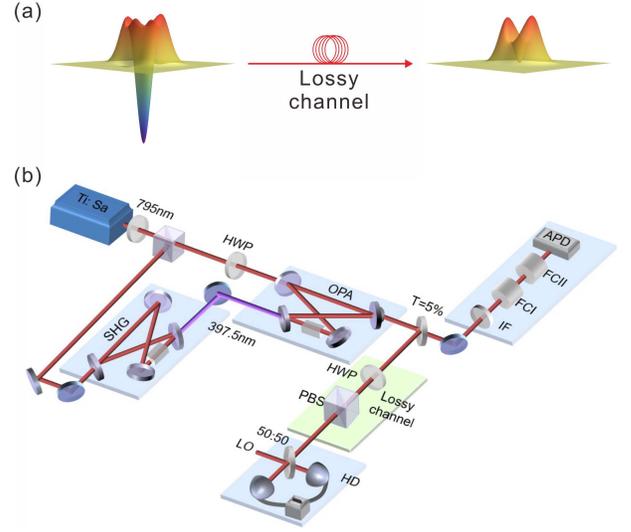}
\end{center}
\caption{(a) Illustration of the decoherence process of optical cat state in phase space. An ideal optical odd cat state with amplitude $\alpha=1$ transmits through a lossy channel with transmission efficiency of 50\%, and the Wigner functions are displayed before and after transmission. (b) Experimental setup. A lossy channel consists of a half wave plate (HWP) and a polarization beam splitter (PBS). OPA, optical parameter amplification with cavity length of $480$ mm; 
SHG, second harmonic generator with cavity length of $480$ mm; 
IF, interference filter ($0.4$ nm); FC, filter cavity with cavity length of 0.75 mm; FC2, filter cavity with cavity length of $2.05$ mm; 
HD, homodyne detector; LO, local oscillator; APD, avalanche photodiode.}
\label{fig:decoherence}
\end{figure}

The schematic of decoherence process of an optical odd cat state in a lossy channel is shown in Fig. \ref{fig:decoherence}(a). An ideal odd cat state has an obvious Wigner function negativity and quantum interference between two coherent states. When it is transmitted through a lossy channel with 50\% transmission efficiency, the Wigner function negativity and quantum interference disappear and the distance of two coherent states is shortened.

When an ideal cat state with amplitude of $\alpha$ is transmitted through a lossy channel with a transmission efficiency of $\eta$, whose density matrix can be expressed as \cite{Glancy2004}
\begin{equation}
\hat{\rho}(\alpha, \eta)=(1-P_{\pm}) \ \hat{\rho}_{\pm}(\sqrt{\eta} \alpha)+ P_{\pm} \ \hat{\rho}_{\mp}(\sqrt{\eta} \alpha),
\end{equation}
where $P_{\pm}=\frac{N_{\mp}(\sqrt{\eta}\alpha)}{2 N_{\pm}(\alpha)}(1-e^{-2(1-\eta)\alpha^{2}})$ is a probability of the initial even (odd) cat state converted to an odd (even) cat state, $\hat{\rho}_{\pm}(\sqrt{\eta} \alpha)=\frac{1}{N_{\pm}(\sqrt{\eta}\alpha)}(| \sqrt{\eta}\alpha \rangle \pm |-\sqrt{\eta}\alpha \rangle)(\langle \sqrt{\eta}\alpha| \pm \langle -\sqrt{\eta}\alpha|)$ is the density matrix of an even (odd) cat state with an amplitude of $\sqrt{\eta}\alpha$ and $N_{\pm}(\alpha)=2(1\pm e^{-2 \alpha^{2}})$ is the normalization factor. According to the density matrix, we can directly calculate the quantum coherence of the cat state in a lossy channel based on relative entropy and $l_{1}$ norm. The relative entropy and $l_{1}$ norm of quantum coherence for the odd cat state are given by
\begin{equation}
C_{rel.ent.}(\alpha, \eta)=(1-P_{-}) C_{rel.ent.}^{odd}(\sqrt{\eta}\alpha)+P_{-} C_{rel.ent.}^{even}(\sqrt{\eta} \alpha), 
\end{equation}
\begin{equation}
C_{l_{1}}(\alpha, \eta)=(1-P_{-})C_{l_{1}}^{odd}(\sqrt{\eta}\alpha) +P_{-}C_{l_{1}}^{even}(\sqrt{\eta}\alpha),
\end{equation}
respectively. The $C_{rel.ent.}^{odd}(\sqrt{\eta}\alpha)$ and $C_{rel.ent.}^{even}(\sqrt{\eta} \alpha)$ are the relative entropy of coherence of the odd cat state and the even cat state with amplitude of $\sqrt{\eta} \alpha$, respectively. The $C_{l_{1}}^{odd}(\sqrt{\eta}\alpha)$ and $C_{l_{1}}^{even}(\sqrt{\eta}\alpha)$ are the $l_{1}$ norm of coherence of the odd cat state and  even cat state with amplitude of $\sqrt{\eta} \alpha$, respectively. The expressions of quantum coherence of the odd and even cat states can be found in Appendix B.

Usually, the fidelity and the Wigner function negativity are used to characterize a cat state. The fidelity of the odd cat state refers to the overlap between the quantum state $\hat \rho$ and an ideal odd cat state $\hat\rho_{-}$ can be defined as $F= \textup{Tr}\ [\hat\rho\  \hat\rho_{-}]$. The fidelity of the odd cat state in a lossy channel is given by
\begin{equation}
F=\cosh(-\alpha^{2}(1-\eta))\frac{\sinh \eta \alpha^{2}}{\sinh \alpha^{2}},
\end{equation}
which is obtained by calculating the overlap between the output state transmitting through the lossy channel and an ideal odd cat state $\hat\rho_{-}(\sqrt{\eta}\alpha)$. The negativity of the Wigner function shows the nonclassical character of quantum state \cite{Tan2020}. It is denoted the minimum of Wigner function $W(x,p)$, where $x$ and $p$ are the position and momentum parameters in phase space, respectively. The negativity of the odd cat state in a lossy channel is denoted as
\begin{equation}
W_{N} = \min \{0, W(0,0)\},
\end{equation}
where $W(0,0)=\frac{1}{\pi N_{-}}e^{-2\alpha^{2}\eta}(2-2e^{-2\alpha
^{2}(1-2\eta)})$.

\begin{figure*}[t]
\begin{center}
\includegraphics[width=150mm]{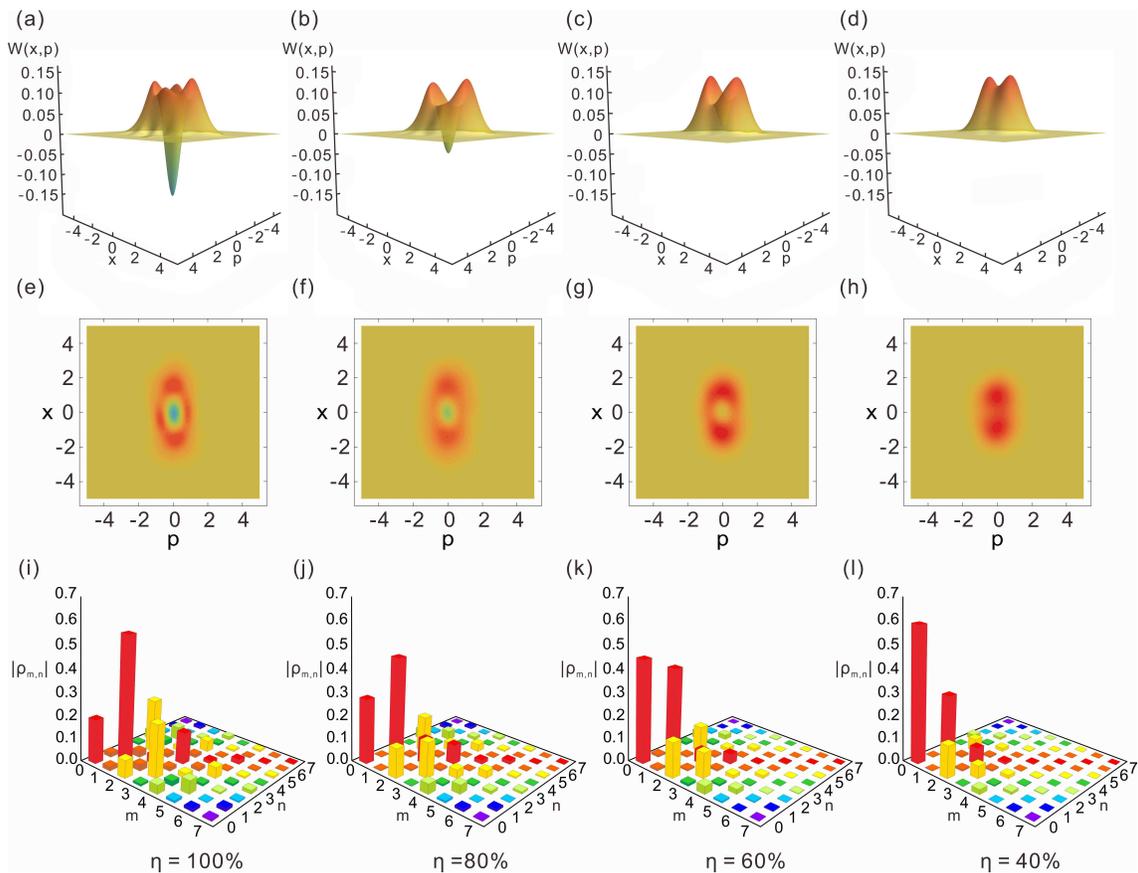}
\end{center}
\caption{Experimental results. The Wigner functions $W(x,p)$ (a-d), projections (e-h) and absolute values of the density matrix elements in the Fock basis (i-l) when the transmission efficiencies are 100\%, 80\%, 60\%, 40\%, respectively. All the results are corrected for 80\% detection efficiency. Only the subspace up to 7 photons is shown.}
\label{fig:matrix}
\end{figure*}

The schematic of our experimental setup is shown in Fig. \ref{fig:decoherence}(b). A continuous wave single frequency Ti: Sapphire laser operated at $795$ nm corresponding to the rubidium D1 line is used as light source. A squeezed vacuum state with $-3\ \textup{dB}$ squeezing is generated by a frequency-degenerate optical parameter amplifier which contains a periodically poled KTiOPO$_{4}$ (PPKTP) crystal with a pump power of 25 mW. By subtracting a photon from the squeezed state through a beam-splitter with a transmissivity of 5\%, an optical cat state is experimentally prepared, which is measured by a homodyne detector when a photon is detected by an avalanche photodiode (APD) in trigger path. The generation rate of the cat state is around $2$ kHz (the maximum dark count is $60$ Hz). We take 50 000 photocurrents to obtain the density matrix in Fock basis with a photon number cutoff of 11 using the iterative maximum-likelihood algorithm \cite{Lvovsky2009} corrected for $80\%$ detection efficiency. The Wigner function of the cat state is reconstructed from the density matrix. The more details of the experimental parameters can be found in Appendix A.

An optical cat state with the amplitude of $1.06\pm 0.01$ and fidelity of $0.68\pm 0.01$ is prepared in our experiment, the corresponding Wigner function, projection of Wigner function and absolute values of the density matrix elements in the Fock basis are shown in Fig. \ref{fig:matrix}(a), \ref{fig:matrix}(e) and \ref{fig:matrix}(i), respectively. From Fig. \ref{fig:matrix}(a), we can see the obvious quantum interference between two coherent states 
and a negative value of Wigner function, which is $-0.16$.

The decoherence of the prepared cat state transmitting through a lossy channel with transmission efficiency $80\%$, $60\%$ and $40\%$ are shown in Fig. \ref{fig:matrix}. We can find the Wigner functions are clearly varied with the loss. The Wigner functions negativity decreases with the decrease of transmission efficiency, and it disappears when the transmission efficiencies are $60\%$ and $40\%$. The phase space distance between the two coherent states ($|\alpha\rangle$ and $|-\alpha\rangle$) becomes smaller with the decrease of transmission efficiency. From Fig. \ref{fig:matrix}(i-l), it is obvious that the probability of Fock state $|1\rangle \langle 1|$ decreases and the probability of vacuum state $|0\rangle \langle0|$ increases with the decrease of transmission efficiency. 

\begin{figure}[tbp]
\begin{center}
\includegraphics[width=\linewidth]{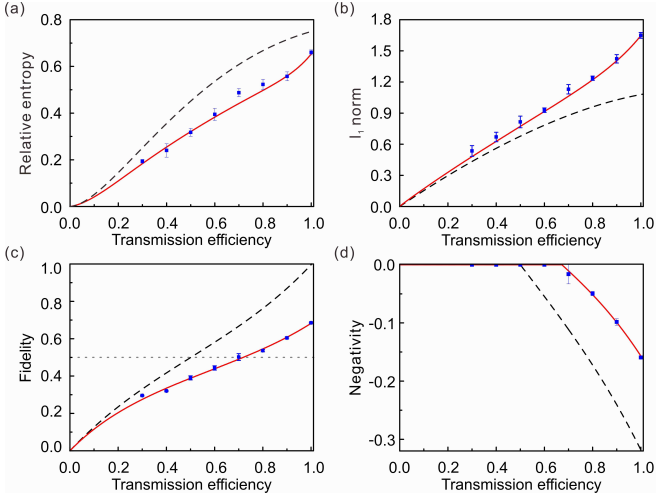}
\end{center}
\caption{Dependence of relative entropy, $l_{1}$ norm, fidelity and negativity of cat states on the transmission efficiency, respectively. The red solid curve and black dashed curve correspond to the experimentally prepared cat state and the ideal cat state, respectively. The blue dots represent the experimental results.}
\label{fig:curve}
\end{figure}

Quantum coherence of the prepared cat state is quantified by relative entropy and $l_{1}$ norm according to Eq. \ref{eq:rel.ent} and Eq. \ref{eq:l1norm}, which are $0.63$ and $1.67$, respectively. The dependence of quantum coherence (including relative entropy and $l_{1}$ norm), fidelity and negativity of Wigner function of the output states on transmission efficiency for the prepared cat state (red solid curve) and an ideal cat state (black dashed curve) with amplitude of 1.06 are shown in Fig. \ref{fig:curve}(a-d), respectively. The relative entropy, $l_{1}$ norm, and fidelity of the cat state all decrease with the decrease of transmission efficiency. The negativity tends to zero with the decrease of transmission efficiency. It is interesting that the coherence still exists when the fidelity is lower than $0.5$ [corresponds to negativity equals to zero for ideal cat states, dotted line in Fig. \ref{fig:curve}(c)] and negativity disappears, which confirms the quantum coherence is robust against loss.

As shown in Fig. \ref{fig:curve}, the theoretical curves of the experimentally prepared cat state and an ideal cat state with the same amplitude are different, which is because the difference between the prepared cat state and an ideal cat state. The curve of relative entropy of coherence of the prepared cat state is lower than that of the ideal cat state. However, the curve of $l_{1}$ norm of coherence of the prepared cat state is higher than that of the ideal cat state. This means that the quantification of quantum coherence of optical cat state is a little bit different using relative entropy and $l_{1}$ norm of coherence. However, the evolutionary trend of quantum coherence of optical cat state quantified by these two means are the same in a lossy channel.


In summary, we experimentally prepare an optical cat state with amplitude of $1.06$ and fidelity of $0.68$ at rubidium D1 line. Then we quantify quantum coherence of the cat state by the relative entropy and $l_{1}$ norm in the Fock basis. The decoherence of quantum coherence, fidelity and Wigner function negativity of the optical cat state in a lossy channel are compared. The experimental results show that quantum coherence of an optical cat state is a robust quantum resource in a lossy environment, since it still exists when the Wigner function negativity of the cat state disappears. We show that the evolution trends of quantum coherence of cat states with relative entropy and $l_{1}$ norm are the same. Our results make a step close to the application based on quantum coherence of cat states.

This research was supported by National Natural Science Foundation of China (11834010); National Key Research and Development Program of China (2016YFA0301402); X. Su thanks the Fund for Shanxi \textquotedblleft 1331 Project \textquotedblright\ Key Subjects Construction.\newline
$^{\dag }$These authors contributed equally to this work.

\appendix	

\section{\label{sec:appendix1}Details of experiment}

A solid-state Titanium-Sapphire continuous wave laser generates $795$ nm light source corresponding to the rubidium D1 line, which is divided into two parts. The first part is injected into a second harmonic generator (SHG) cavity with cavity length of $480$ mm, which contains two concave mirrors (R=$50$ mm), a high reflectivity mirror, a plane mirror with the transmissivity of $8 \%$, and a periodically poled KTiOPO$_{4}$ (PPKTP, $1\times 2\times 10$ mm$^{3}$) crystal. The second part passes through the model cleaner and is divided into local oscillator of homodyne detector and signal beam of a frequency degenerate optical parameter amplifier (OPA), respectively. The signal beam is periodically chopped by two acousto-optic modulators (AOMs), which defines distinct time bins by presence and absence of the signal beam, and thus enables the alternate sequence of locking and photon counting without the signal beam. When the signal beam is absented, we generate a nearly pure squeezed vacuum state with $-3\  \textup{dB}$ squeezing by an OPA with cavity length of $480$ mm, which contains two concave mirrors (R=$50$ mm), a high reflectivity mirror, a plane mirror with the transmissivity of $12.5 \%$ and a PPKTP crystal. The OPA works on the parametric amplification status in our experiment, which is guaranteed by locking the relative phase between pump beam and signal beam to zero.

A beam-splitter, which is composed by a half wave plate and a polarization beam splitter (PBS), transmits 5\% of the squeezed vacuum state toward an avalanche photodiode (APD) through a filtering system for photon counting. The filter system consists of an interference filter (0.4 nm) and two filter cavities with fineness of $1200$, whose cavity lengths are 0.75 mm and 2.05 mm respectively.

The reflected beam by the beam-splitter is measured by a homodyne detector when a photon is detected by the APD. The bandwidth of the homodyne detector is 30 MHz and the detection efficiency is about 80\%, which includes four parts: the interference efficiency between signal light 
and local light (98.5\%), quantum efficiency of photodiodes (92\%, S3883), the 19 dB clearance with the 16 mW local oscillator at 13 MHz of the homodyne detector (corresponding to equivalent efficiency 98.7\%), and transmission efficiency (91\%). Using the maximum likelihood algorithm, we obtain the density matrix and the associated Wigner function.

\section{Quantum coherence of cat states} 

Optical cat states can be expressed as a quantum superposition of two coherent states
\begin{equation}
|\Psi \rangle =\frac{1}{\sqrt{N\pm}}(|\alpha \rangle \pm |-\alpha
\rangle),
\end{equation}
where $N_{\pm}=2(1\pm e^{-2|\alpha|^{2}})$ are the normalization factors, $+$ and $-$ correspond to the even and odd cat states, respectively. Even and odd cat states can be expressed in Fock basis as \cite{Neergaard-Nielsen}
\begin{equation}
|\Psi_{+} \rangle =\frac{1}{\sqrt{\cosh |\alpha|^{2}}} \sum_{n=0}^{\infty }\frac{\alpha^{n}}{\sqrt{\ n! \ }}\delta_{0n}^{[2]}\ |n\rangle,
\end{equation}
\begin{equation}
|\Psi_{-} \rangle =\frac{1}{\sqrt{\sinh|\alpha|^{2}}} \sum_{n=0}^{\infty }\frac{\alpha^{n}}{\sqrt{\ n!\ }}\delta_{1n}^{[2]}\ |n\rangle,
\end{equation}
respectively. $\delta_{0n}^{[2]}=0$, $\delta_{1n}^{[2]}=1$ when $n$ is odd and $\delta_{0n}^{[2]}=1$, $\delta_{1n}^{[2]}=0$ when $n$ is even. 

Although an ideal cat state is expressed in the infinite dimensional Fock basis, it can be expressed in the finite dimension when the density matrix can characterize the information of the cat state. For example, the odd cat state with amplitude of 1.3 is represented by a density matrix with 11 dimension \cite{Neergaard-Nielsen2006}, and the even cat state with amplitude of 1.61 is represented by a density matrix with 10 dimension \cite{Takahashi2008}. For an odd cat state with amplitude of $\alpha$, the density matrix elements are expressed as $\frac{1}{\sinh|\alpha|^{2}} \frac{\alpha^{m+n}}{\sqrt{ m! n!}}\delta_{1m}^{[2]}\delta_{1n}^{[2]}$, and the probability of the $p_{|n \rangle \langle n|}$ is $\frac{1}{\sinh|\alpha|^{2}} \frac{\alpha^{2n}}{n!}\delta_{1n}^{[2]}$. In the experiment, it is truncated to the dimension $d$ (corresponding to a photon-number cutoff of $d-1$) when the probability of higher photons ($d, ... , \infty$ ) is small enough to be ignored (e. g. $10^{-3}$).

Considering $d$-dimensional Hilbert space, the relative entropy of coherence of the even and odd cat states can be expressed as
\begin{eqnarray}
C_{rel.ent.}^{even} (\alpha)=& &\frac{1}{\cosh |\alpha|^{2}}(\sum_{n=0}^{d-1}{\frac{\alpha^{2n}}{n!}\delta_{0n}^{[2]}} \log_{2} \cosh |\alpha|^{2}
\nonumber \\ & & 
-\sum_{n=0}^{d-1}{\frac{\alpha^{2n}}{n!}\delta_{0n}^{[2]}}\log_{2} \frac{\alpha^{2n}}{ n!}),
\end{eqnarray}
\begin{eqnarray}
C_{rel.ent.}^{odd}(\alpha)=& &\frac{1}{\sinh |\alpha|^{2}}(\sum_{n=0}^{d-1}{\frac{\alpha^{2n}}{n!}\delta_{1n}^{[2]}} \log_{2} \sinh|\alpha|^{2}
\nonumber \\ & &
-\sum_{n=0}^{d-1}{\frac{\alpha^{2n}}{n!}\delta_{1n}^{[2]}}\log_{2} \frac{\alpha^{2n}}{ n!}).
\end{eqnarray}

The $l_{1}$ norm of coherence of the even and odd cat states can be expressed as
\begin{equation}
C_{l_{1}}^{even}(\alpha)=\frac{1}{\cosh |\alpha|^{2}} \sum_{\substack{m,n=0\\ m\neq n}}^{d-1} {\frac{\alpha^{m+n}}{\sqrt{m\,!\, n\,!}}\ \delta_{0m}^{[2]}\ \delta_{0n}^{[2]}},
\end{equation}
\begin{equation}
C_{l_{1}}^{odd}(\alpha)=\frac{1}{\sinh |\alpha|^{2}} \sum_{\substack{m,n=0\\ m\neq n}}^{d-1} {\frac{\alpha^{m+n}}{\sqrt{m\,!\, n\,!}}\ \delta_{1m}^{[2]}\ \delta_{1n}^{[2]}}.
\end{equation}

\begin{figure}[h]
\begin{center}
\includegraphics[width=\linewidth]{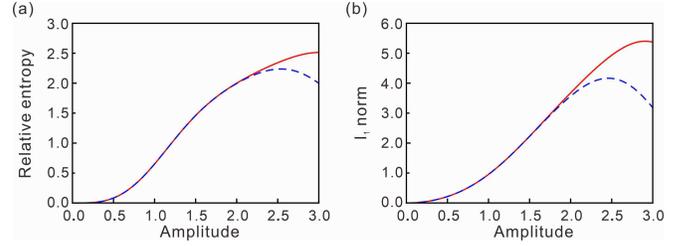}
\end{center}
\caption{Dependence of relative entropy and $l_{1}$ norm of quantum coherence of an optical cat state on the amplitude of it, respectively. The red solid and blue dashed curves correspond to the results in 16-dimensional and 12-dimensional Hilbert space, respectively.}
\label{fig:dimension}
\end{figure}

The dependence of relative entropy and $l_{1}$ norm of quantum coherence of an odd cat state on the amplitude of it with different dimensions are shown in Fig. \ref{fig:dimension}. We can see when the amplitude is less than 2, the quantum coherence of cat states in 16 or 12-dimensional Hilbert space are the same, which indicates the 12-dimensional Hilbert space we used in our experiment is enough to estimate the quantum coherence of cat states with small amplitudes.

\end{document}